\documentclass[12pt]{article}
\usepackage{amssymb,amsmath, amsthm}
\usepackage{ifthen, xcolor}
\makeatletter

\@addtoreset{equation}{section}
\def\section{\@startsection {section}{1}{\z@}{-2.5ex plus -1ex minus
 -.2ex}{1.3ex plus .2ex}{\large\bf}}
\def\subsection{\@startsection{subsection}{2}{\z@}{-2.25ex plus%
 -1ex minus -.2ex}{0.5ex plus .2ex}{\bf}}
\advance \voffset by -0.8in
\advance \hoffset by -0.6in
\def\aa{a}
\def\ha{\alpha}
\def\bgp{\g{P}^\circ}

\def\NW{\hat{G}_0}
\def\NWH{G_0}
\def\NHH{NH}

\def\newS{\tilde{S}}
\def\newJ{\tilde{J}}
\def\newM{\tilde{M}}
\def\newH{\tilde{H}}
\def\INW{\hat{G}}
\def\nawi{\hat{\mathfrak{g}}_0}
\def\heis{\hat{\mathfrak{h}}_1}
\def\HEIS{\hat{H}_1}
\def\nh{\mathfrak{nh}}
\def\manif{\mathcal M}
\def\inawi{\hat{\mathfrak{g}}}

\def\sig{\zeta}

\def\nh{\hat{\mathfrak{nh}}}

\def\La{\Lambda}

\def\NH{\hat{NH}}

\def\bpm{\begin{pmatrix}}
\def\epm{\end{pmatrix}}

\newcommand{\RR}{\mathbb{R}}
\newcommand{\CC}{\mathbb{C}}

\def\bea{\begin{eqnarray}}
\def\eea{\end{eqnarray}}
\theoremstyle{definition}
 \newtheorem{theorem}{Theorem}[section]
\newtheorem{lemma}[theorem]{Lemma}

\DeclareMathOperator{\ad}{ad}

\newcommand{\lie}[2]{\langle #1\,,\,#2\rangle}

\newcommand{\comm}[2]{\bigl[ #1\,,\,#2\bigr]}

\newcommand{\dd}{\mathrm{d}}
\newcommand{\ii}{\mathrm{i}}

\newcommand{\g}[1]%
{%
\ifthenelse{\equal{#1}{a}}{\alpha}{}%
\ifthenelse{\equal{#1}{b}}{\beta}{}%
\ifthenelse{\equal{#1}{g}}{\gamma}{}%
\ifthenelse{\equal{#1}{G}}{\Gamma}{}%
\ifthenelse{\equal{#1}{d}}{\delta}{}%
\ifthenelse{\equal{#1}{D}}{\Delta}{}%
\ifthenelse{\equal{#1}{e}}{\epsilon}{}%
\ifthenelse{\equal{#1}{z}}{\zeta}{}%
\ifthenelse{\equal{#1}{h}}{\eta}{}%
\ifthenelse{\equal{#1}{8}}{\theta}{}%
\ifthenelse{\equal{#1}{i}}{\iota}{}%
\ifthenelse{\equal{#1}{k}}{\kappa}{}%
\ifthenelse{\equal{#1}{l}}{\lambda}{}%
\ifthenelse{\equal{#1}{L}}{\Lambda}{}%
\ifthenelse{\equal{#1}{m}}{\mu}{}%
\ifthenelse{\equal{#1}{n}}{\nu}{}%
\ifthenelse{\equal{#1}{ks}}{\xi}{}%
\ifthenelse{\equal{#1}{p}}{\pi}{}%
\ifthenelse{\equal{#1}{P}}{\Pi}{}%
\ifthenelse{\equal{#1}{r}}{\rho}{}%
\ifthenelse{\equal{#1}{s}}{\sigma}{}%
\ifthenelse{\equal{#1}{t}}{\tau}{}%
\ifthenelse{\equal{#1}{y}}{\ypsilon}{}%
\ifthenelse{\equal{#1}{f}}{\phi}{}%
\ifthenelse{\equal{#1}{x}}{\chi}{}%
\ifthenelse{\equal{#1}{ps}}{\psi}{}%
\ifthenelse{\equal{#1}{Ps}}{\Psi}{}%
\ifthenelse{\equal{#1}{w}}{\omega}{}%
\ifthenelse{\equal{#1}{W}}{\Omega}{}%
\ifthenelse{\equal{#1}{ff}}{\varphi}{}%
\ifthenelse{\equal{#1}{ee}}{\varepsilon}{}%
}

\begin{document}
\parskip 4pt
\parindent 10pt
\begin{flushright}
EMPG-10-09\\
\end{flushright}

\begin{center}
\baselineskip 24 pt 
{\Large \bf   Galilean quantum gravity with cosmological constant and the  extended $q$-Heisenberg algebra}

\baselineskip 16 pt

\vspace{.5cm}
G.~Papageorgiou{\footnote{\tt georgios@ma.hw.ac.uk} and { B.~J.~Schroers}\footnote{\tt bernd@ma.hw.ac.uk} \\
Department of Mathematics and Maxwell Institute for Mathematical Sciences \\
 Heriot-Watt University \\
Edinburgh EH14 4AS, United Kingdom } \\

\vspace{0.5cm}

{1 August 2010}

\end{center}

\begin{abstract}
\noindent
We  define a theory of  Galilean gravity in 2+1 dimensions with cosmological constant as a Chern-Simons gauge theory of  the  doubly-extended Newton-Hooke group, extending our previous study  of classical and quantum gravity in 2+1 dimensions in the Galilean limit.  We exhibit an $r$-matrix which is compatible with  our Chern-Simons action (in a sense to be defined) and show that the associated bi-algebra structure of  the Newton-Hooke Lie algebra is that of the classical double of the extended Heisenberg algebra.  We deduce that, in the quantisation of the theory according to the combinatorial quantisation programme, much of the quantum theory is determined by the quantum double of the extended $q$-deformed Heisenberg algebra.  
\end{abstract}


\section{Introduction}
The purpose of this paper is to show how one can include a non-vanishing cosmological constant in the Galilean limit of classical and quantum gravity in 2+1 dimensions. The framework for our discussion is provided by our extended treatment  of Galilean gravity in 2+1 dimensions in our previous paper \cite{Papageorgiou:2009zc}. Our approach in writing  the current paper is to refer to that paper for all details about Galilean (quantum) gravity in 2+1 dimensions,  and to focus on those aspects of the treatment in \cite{Papageorgiou:2009zc} which are changed by the presence of a cosmological constant.  Our main conclusion is that quantum double of the extended $q$-deformed Heisenberg algebra plays the role of a `quantum isometry group' in  (2+1)-dimensional Galilean quantum gravity with cosmological constant (which fixes $q$). Exploring the structure of this double and its application to  Galilean quantum gravity in detail is beyond the scope of this paper, and left for future work.

Over the last decade it has become clear that three spacetime dimensions provide a setting where the relation between quantum gravity on the one hand and non-commutative geometry and quantum groups on the other  can be investigated concretely and in a  mathematically satisfactory fashion, see e.g. \cite{BM,Schroers,BNR,MS2,Freidel:2005me,schroers-2007-035,Joung:2008mr,MS}. Remarkably,  the picture that has emerged shows many of the most studied quantum groups (e.g. $U_q(sl_2)$ and $U_q(su_2)$) and constructions (e.g. the quantum double and bicrossproduct construction)  arising naturally in three-dimensional quantum gravity. A secondary purpose of the current paper is thus to add  a further detail to this picture by  
placing yet another popular quantum group - the extended $q$-Heisenberg or oscillator algebra - in the context of  three-dimensional quantum gravity.

We have  endeavoured to explain our  results without duplicating our detailed discussion in \cite{Papageorgiou:2009zc} of Galilean quantum gravity with  vanishing cosmological constant. The remainder of this introduction is thus a short summary of the  approach and key results of \cite{Papageorgiou:2009zc}. 

One of the problems motivating   the paper \cite{Papageorgiou:2009zc} is that  the formulation of (relativistic) gravity  in 2+1 dimensions as a Chern-Simons gauge theory of the Poincar\'e group does not have a good Galilean limit because  the invariant inner product on the Poincar\'e Lie algebra, which is essential for the Chern-Simons formulation, degenerates in the  limit $c\rightarrow 0$. However, if one considers  a trivial two-fold extension of the Poincar\'e Lie algebra  and then takes the limit $c\rightarrow 0$ it is possible to obtain a non-trivial two-fold extension of the Galilei Lie algebra which does have an invariant inner product and thus allows for a  Chern-Simons formulation of Galilean gravity. The two central generators in  the extended Galilei group also  turn out to be required  for describing the mass and spin of particles coupled to gravity. Most importantly, the Lie algebra of the  doubly extended  Galilei group has the structure of a Lie bi-algebra with a classical $r$-matrix which is compatible with the Chern-Simons action in the sense of Fock and Rosly \cite{FR}:  its symmetric part is equal to the quadratic Casimir associated to the inner product used in the definition of the Chern-Simons action.  This means that one can use the framework of Fock and Rosly to describe the Poisson structure on the phase space of the Chern-Simons theory and that one can quantise the theory using the framework of Hamiltonian or combinatorial  quantisation \cite{AGSI,AGSII,AS}.

 The geometrical interpretation  of the Chern-Simons formulation and combinatorial quantisation for (relativistic) gravity is reviewed,  for example, in 
 \cite{schroers-2007-035,MS} and can be summarised as follows. The gauge group used in the Chern-Simons action is the isometry group of a model spacetime. The model spacetime depends on the signature (Lorentzian or Euclidean) and on the cosmological constant. For example, Minkowski space is the model spacetime for Lorentzian gravity with vanishing cosmological constant.  Classical solutions of the Chern-Simons equations of motion are flat connections and describe universes which are locally isometric to the model spacetime but may be patched together to give a globally non-trivial solution. Particles can be coupled to the theory by minimal coupling between the gauge field and co-adjoint orbits of the gauge group. Geometrically, they act as defects in the model spacetimes, with a particle's mass determining a wedge disclination and a  particle's spin causing a screw dislocation. The phase space of Chern-Simons theory coupled to a fixed number of particles parametrises  spacetime geometries of this form. Following Fock and Rosly \cite{FR},   the Poisson structure on the phase space can be expressed in terms of a  compatible $r$-matrix.    In the quantum theory, the geometrical picture gets deformed, with the role of the isometry group now being played by a quantum group which quantises the Lie bi-algebra determined by the classical $r$-matrix. It seems likely that the resulting picture in the quantum theory  can be described  geometrically in analogy with the classical situation, but using non-commutative instead of commutative geometry. The details of this have not been worked out, but the construction of the Hilbert space in terms of the representation theory of the quantum isometry group is  known, see again \cite{MS} for a summary.

In this paper we give a Chern-Simons formulation of Galilean gravity in 2+1 dimensions with a cosmological constant and identify the quantum group which controls the combinatorial quantisation in this case. 
We have  organised our account  as follows. In Sect.~2 we define the doubly extended Newton-Hooke Lie algebra and group,  introduce Newton-Hooke spacetimes as hypersurfaces in an auxiliary four-dimensional space, show how they can alternatively be realised  as cosets of the (unextended) Newton-Hooke group and deduce how the Newton-Hooke group acts on Newton-Hooke spacetimes. We briefly discuss  inertial motions  and show that the trajectories in the spatial plane are  ellipses, straight lines or hyperbolae if the  Galilean cosmological constant is, respectively, negative,  zero or positive.  
In Sect.~3 we give  the Chern-Simons formulation  of Galilean gravity with cosmological constant in 2+1 dimensions and recall, very briefly, the quantisation of a Chern-Simons theory via the combinatorial or Hamiltonian approach. 
Sect.~4 contains an account of the bi-algebra structure of  the extended Heisenberg algebra and of the doubly extended Newton-Hooke Lie algebra as the  classical double of the extended Heisenberg algebra. We   exhibit the associated $r$-matrix 
of the Newton-Hooke bi-algebra and  
and show that it is compatible with the Chern-Simons action of Sect.~3 in the sense of Fock and Rosly. In Sect.~5 we 
briefly review the quantisation of the extended Heisenberg algebra to the extended $q$-Heisenberg algebra and discuss the $*$-structures which are relevant in our context. 
Following the general principle that a quantisation of a classical double of a  given Lie bi-algebra is provided  by the  quantum double of the Hopf algebra which quantises the given Lie bi-algebra we deduce, in our final Sect.~5,  that the quantum double of the  extended $q$-deformed Heisenberg algebra provides a quantisation of Newton-Hooke bi-algebra structure of Sect.~4. We postpone a detailed investigation of this double and its representation   but  discuss  how the results of such an investigation would relate to Galilean quantum gravity with a cosmological constant. 

\section{The Newton-Hooke group and associated  classical spacetimes}

\subsection{Lie algebra and group structure}
\label{sec:newton-hooke-groups}

We use the notation $ \nh $  for  the Lie algebra of a two-fold central extension of  the Newton-Hooke group in 2+1 dimensions which is sometimes called `exotic' Newton-Hooke symmetry \cite{Alvarez:2007fw}. As we shall see below, the structure of $\nh$  as a  real Lie algebra really depends on the sign of a parameter $\lambda$. In the literature on this subject the two real algebras are usually distinguished by writing $\nh_+$ and $\nh_-$. However, in our treatment we have endeavoured to give, as far as possible,  a unified treatment of the two  signs. We therefore use the notation $\nh$ for both signs, and specify the sign of $\lambda$ when necessary.

 The Lie algebra $\nh$ is  eight dimensional, and a possible basis consists of a  rotation generator $J$, two (Galilean) boost generators $J_i$, $i=1,2$, a time translation generator $H$, two spatial translation generators $P_i$, $i=1,2$, as well as the central elements $S$ and $M$ which, in the particle interpretation of the irreducible representations,  represent the rest spin and rest mass. The Lie algebra can be obtained via contraction from trivial extensions of the de Sitter and anti-de Sitter groups in 2+1 dimensions. There is considerable literature on Newton-Hooke symmetries and spacetimes both in 3+1 and 2+1 dimensions,  starting with the classic paper \cite{BLL} and continuing with recent studies of the (2+1)-dimensional \cite{arratia-1999, Alvarez:2007fw}  and (3+1)-dimensional
\cite{Mech_NH, NH_spacetimes_Gibbons} situation.  Further details in the notation used here can be found in \cite{Papageorgiou:2009zc}. The extended Newton-Hooke Lie algebra depends on a constant $\lambda$ which is the {\em negative} Galilean cosmological constant in the sense that is obtained from the usual relativistic cosmological constant $\Lambda$ by  setting 
\begin{equation}
\label{cosdef}
\lambda =-c^2\Lambda
\end{equation} 
and taking the limit $c\rightarrow \infty$ and $\Lambda\rightarrow 0$ in such a way that $\lambda$  remains finite.  
 The brackets of the extended Newton-Hooke Lie algebra  are 
\begin{equation}
 \begin{aligned}
 \label{NewtonHookeimproved}
 [J_i,J_j]&=\epsilon_{ij}S & [J_i,J]=&\epsilon_{ij}J_j &[S,\cdot]=[M,\cdot]=&0\\
 [J_i,P_j]&=\g{e}_{ij} M, & [J_i,H]=&\epsilon_{ij}P_j,&[P_i,J]=&\epsilon_{ij}P_j \\
  [P_i,P_j]&=\epsilon_{ij}\lambda S, & [P_i,H]=&\epsilon_{ij}\lambda J_j&[J, H]=&0.
   \end{aligned}
\end{equation}
In the limit $ \g{l}\rightarrow 0 $  we obtain
  a  two-fold central extension of the  Galilei Lie  algebra, which we
  denote $\inawi$.  In our previous work \cite{Papageorgiou:2009zc} we
  studied the Chern-Simons theory with a  gauge group $ \INW $ whose Lie
  algebra is $ \inawi $ and interpreted it as the theory of Galilean
  gravity in 2+1 dimensions.  The doubly extended Galilei group in 2+1 dimensions is of interest in relation to planar physics and non-relativistic anyons. Many of its interesting features (see, for example,  \cite{LSZ,DH, Jackiw:2000tz, HP1}) presumably persist in the presence of a  cosmological parameter $\lambda$, but we will not pursue this here. 
   
  Since keeping track of physical dimensions is important  at various points in this paper we  note  here  that the rotation generator $J$ is dimensionless, the boost generators $J_i$ have the dimension of inverse velocity and
the central element $S$ has the dimension of inverse velocity squared. The Hamiltonian $H$ has the dimension of inverse time, the translation generators $P_i$ have the dimension of inverse length, and the second central element  $M$  has the dimension of time divided by length squared. As usual, multiplying these dimensions by the dimension of an action gives the dimension of the associated observable in the quantum theory:  angular momentum for $J$, momentum for $P_i$, energy for $H$, mass for $M$ and so on.  Note also that, with the usual, relativistic cosmological constant $\Lambda$ having the dimension of inverse length squared, the dimension of $\lambda$ is inverse time squared. 

The subalgebra  $\nawi$  spanned by $\{J,S,J_1,J_2\}$
 with non-zero Lie brackets
 \begin{equation}
    \label{nawij}
   [J_1,J_2]= S \quad [J_i,J]=\epsilon_{ij}J_j  
   \end{equation}
  is  a central extension of the Lie
   algebra of the Euclidean group in two dimensions\footnote{We will
use hatted symbols to denote centrally extended algebraic objects and
the same symbols without the hat to denote the same objects without
their extensions}. In  the string
   theory literature it is sometimes called the Nappi-Witten Lie
   algebra \cite{Nappi:1993ie,FoFS}.  It can also be viewed as the
   Heisenberg algebra with an outer automorphism or as the  harmonic oscillator Lie algebra. To make this manifest we
carry out a complex basis  change to 
\begin{equation}
   \newS = -2\ii S,  \quad \newJ = \ii J, \quad  Z = J_1 + \ii J_2, \quad 
\bar{Z} =  J_1 - \ii J_2. 
\end{equation}
In this basis,  the brackets  \eqref{nawij} are  
\begin{align}
  \label{eq:nawi_z}
  \comm{Z}{\bar{Z}} =& \newS & \comm{Z}{\newJ} =& Z &
  \comm{\bar{Z}}{\newJ} =& - \bar{Z},
\end{align}
showing that, in the harmonic oscillator interpretation,  $\newJ$ plays the role the  number operator and $\bar Z$  and $Z$  play the role of raising and lowering operators.  Note that
the structure constants are real both in \eqref{nawij} and \eqref{eq:nawi_z}, showing that
these are different real forms of the complexified Lie algebra  $\nawi\otimes \CC$. We will distinguish the 
real forms by using the notation $\nawi$ for the algebra \eqref{nawij} and $\heis$ for the algebra \eqref{eq:nawi_z}.
For later use we note that a  faithful  $ 3 \times 3 $ matrix representation of 
$\heis$ is given by
\begin{equation}
\begin{aligned}
  \label{eq:matrices}
  \rho(\newJ)=& \begin{pmatrix}
 0 & 0 & 0 \\
 0 & 1 & 0 \\
 0 & 0 & 0
\end{pmatrix}, &&  \rho( Z) =&
\begin{pmatrix}
 0 & 1 & 0 \\
 0 & 0 & 0 \\
 0 & 0 & 0
\end{pmatrix}, \\
\rho(\bar{Z}) =& 
\begin{pmatrix}
 0 & 0 & 0 \\
 0 & 0 & 1 \\
 0 & 0 & 0
\end{pmatrix},&&
\rho(\newS)=& 
\begin{pmatrix}
 0 & 0 & 1 \\
 0 & 0 & 0 \\
 0 & 0 & 0
\end{pmatrix}.
\end{aligned}
\end{equation}

The  full Lie algebra $\nh$ can be viewed as   a generalised complexification   of the homogeneous subalgebra $\nawi$ in the sense that it can be obtained by tensoring the Lie 
algebra  $\nawi$ with a ring $ \mathcal{R}_{\g{l}} $ of numbers  of the form $a+\theta b$, with $a,b\in\RR$ and
the formal parameter $\theta$ satisfying  $\theta^2=\lambda$.  This approach
has proved  useful    in the context of
(relativistic) three-dimensional gravity. It was introduced for
vanishing cosmological constant in \cite{Martin}, generalised to
arbitrary values of the  cosmological constant in  \cite{Meusburger}
and further developed in \cite{MS6}. A formal definition is given in \cite{Meusburger}.
 In this formulation we recover  the brackets given in \eqref{NewtonHookeimproved} from \eqref{nawij} by setting $H=\theta J,
P_i=\theta J_i$ and $M=\theta S$.

Next, we turn to the Lie groups $\NW$ and  $\NH$ whose Lie algebras are, respectively, 
 $ \nawi$ and $\nh$.  As in \cite{Papageorgiou:2009zc} we write elements  
of $ \NW $ as tuples
\begin{equation}
  \label{eq:nwgroup}
  (\varphi,\vec{w},\sig) \leftrightarrow \exp(\varphi J) \exp(\vec{w}
  \vec{J}) \exp(\sig S).
\end{equation}
The composition law  can be written as 
   \begin{equation}
   \label{nawimulti}
   (\varphi_1,\vec{w}_1,\sig_1) (\varphi_2,\vec{w}_2,\sig_2)=(\varphi_1+\varphi_2,
   \vec{w}_1 + R(\varphi_1)\vec{w}_2,\sig),
   \end{equation}
   where 
   \begin{equation}
   \label{rot}
     R(\varphi) = \begin{pmatrix} \cos\varphi & \sin\varphi \\-\sin\varphi & \cos\varphi \end{pmatrix} 
     \end{equation}
  is the  $ SO(2) $ matrix implementing a rotation of the plane by $\varphi$,  and 
   \begin{equation}
   \label{sigmult}
   \sig = \sig_1+\sig_2 +\frac 1 2 \vec{w}_1\times
   R(\varphi_1)\vec{w}_2. 
   \end{equation} 
   Thus, we have the group structure 
   \begin{equation}
   \label{nwiso}
   \NW \simeq (SO(2)\ltimes \RR^2)\ltimes \RR.
  \end{equation}

One  checks that, in the matrix representation \eqref{eq:matrices} of $\heis$,  
\begin{align}
  \label{eq:matrixnawi}
 \rho( e^{z\newS }e^{x\bar{Z}} e^{yZ}e^{\phi \newJ})=\left(
\begin{array}{lll}
 1 & e^{\phi } y & z+ x y \\
 0 & e^{\phi } & x \\
 0 & 0 & 1
\end{array}
\right),
\end{align}
for real parameters $\phi, x,y, z$.  The group of all such matrices is the Heisenberg group with an outer autormophism,  and we denote it by $\HEIS$.

The  group law for $ \NH $ can be obtained by a generalised  complexification of 
the parameters in the above result i.e. by writing  elements $ g \in \NH $
as
\begin{equation}
  \label{complmulti}
  (\varphi + \g{8}\g{a},\vec{w} +  \g{8}\vec{a}, \sig + \g{8} \g{h}).
\end{equation}
The group law can then be extracted from \eqref{nawimulti} and  \eqref{sigmult}. Details of how this works for the case $\lambda =0$ are given in \cite{Papageorgiou:2009zc}.\\

Working with the `complexified' notation is  efficient for many calculations and allows one to treat the different signs of the cosmological constant  in a unified fashion. It is nonetheless worth noting the different structures of the 
Newton-Hooke group for   different values of $ \g{l} $.  In the case $\lambda =0$ we obtain a semi-direct product structure which was discussed  in \cite{Papageorgiou:2009zc}.
For $ \g{l} > 0  $  we can define the generators
\begin{equation}
  \label{eq:NH_two_copies}
  \begin{aligned}
  J^+ &= \dfrac{1}{2}(J + \dfrac{1}{\sqrt{\g{l}}}H)      
  &&& J^- &= \frac{1}{2}(J
  - \dfrac{1}{\sqrt{\g{l}}} H), \\
  \vec{J}^+ &= \dfrac{1}{2}(\vec{J} + \dfrac{1}{\sqrt{\g{l}}} \vec{P})   
  &&&\vec{J}^-&= \dfrac{1}{2} (\vec{J} -
    \dfrac{1}{\sqrt{\g{l}}} \vec{P}), \\
  S^+ &= \dfrac{1}{2} (S + \dfrac{1}{\sqrt{\g{l}}} M)
  &&& S^- &= \dfrac{1}{2}(S -
  \dfrac{1}{\sqrt{\g{l}}}M),
  \end{aligned}
\end{equation}
and check that the algebras 
\[
\mathfrak{g}^+ = \{ J^+, J_1^+,J^+_2,  S^+ \}, 
\quad 
\mathfrak{g}^-  = \{ J^-,  J^-_1, J^-_2,  S^-  \}
\]
 each satisfy the commutation
relations of the extended homogeneous Galilei algebra
(\ref{nawij}).  It is also  straightforward to show that 
$[\mathfrak{g}^+, \mathfrak{g}^-] = 0 $. Thus, for $\lambda >0$, $ \mathfrak{nh}$ has the direct sum structure $\mathfrak{g}^+ \oplus \mathfrak{g}^-$. 
At the group level we have $ \NH \cong \NW \times \NW $ for $\lambda >0$. 

When $\lambda <0$  the Newton-Hooke group is a 
complexification of the homogeneous
group $\NW$ in the usual  sense. Indeed, setting $ \g{8}= \ii \sqrt{- \g{l}}
$,    looking at the  group law \eqref{nawimulti} and inserting the parametrisation \eqref{complmulti} 
we deduce that, when $\lambda <0$,  $ \NH \cong (\CC^*\ltimes \CC^2)\ltimes
\CC $, where $\CC^*$ is the multiplicative group of non-zero complex numbers, $\CC^2$ and $\CC$ are viewed as additive groups and the group composition law can be read off from \eqref{nawimulti} for complex parameters. 

\subsection{Newton-Hooke spacetimes as  hypersurfaces}

We now turn to the spacetimes on which the symmetry groups of the previous sections act. We shall see that the Newton-Hooke groups act  on three-dimensional spacetimes, called Newton-Hooke spacetimes, which  can be naturally realised as hypersurfaces in a four-dimensional ambient space. The three-dimensional spacetimes are equipped with two structures - an absolute time and a spatial euclidean structure - which are  invariant under the Newton-Hooke group action and which are induced from the embedding in the ambient four-dimensional space.

In order to construct the Newton-Hooke spacetimes we begin with the model spacetime of relativistic three-dimensional gravity, reviewed, for example in \cite{Papageorgiou:2009zc}. These can be realised as hypersurfaces in  $\RR^4$ equipped with a metric
\begin{equation}
\label{met}
g_{\mu\nu}=\text{diag}\left(- {c^2},1,1,\frac{1}{\g{L}}\right),
\end{equation}
 which depends on the cosmological constant  $\Lambda$ and the speed of light  $c$.  Explicitly, we define the two-parameter family of 
three-dimensional hypersurfaces
\begin{equation} 
\label{hypersur}
H_{c,\La}=
\left\{(t,x,y, w)\in \RR^4 | -c^2t^2 + x^2 + y^2 + \frac {1}{\La} w^2 = \frac{1}{\La}\right\}.
\end{equation}
  As explained in \cite{Papageorgiou:2009zc},   Euclidean model spacetimes like the three-sphere and hyperbolic three-space can be included in this family by allowing  $c^2$ to take negative values.  For all values  of  $\Lambda$ and $c\neq \infty$, the local isometry groups of  three-dimensional gravity can be recovered as group of the linear transformations of $\RR^4$ which leave the metric \eqref{met} (and hence the hypersurfaces \eqref{hypersur}) invariant. 

In order to take the Galilean limit of  $H_{c,\La}$ we again use the definition \eqref{cosdef} and   re-write the defining equation of the hypersurfaces as follows
\begin{equation} 
\label{hypersurr}
H_{c,\La}=
\left\{(t,x,y, w)\in \RR^4 | w^2+\lambda t^2 + \Lambda(x^2 + y^2)  = 1 \right\}.
\end{equation}
Then we take the limit $c\rightarrow \infty $ in such a way that $\lambda$  remains finite. This necessarily requires $\Lambda \rightarrow 0$, so that the equation defining  Newton-Hooke spacetimes becomes a condition on $t$ and $w$ only:
\begin{equation} 
\label{hypersurg}
H_{\lambda}=
\left\{(t,x,y, w)\in \RR^4 | w^2+\lambda t^2  = 1 \right\}.
\end{equation}
Geometrically, Newton-Hooke spacetimes are products of a one dimensional manifold $N$ which parametrises time and  the spatial plane $\RR^2$, embedded in $\RR^4$ according to \eqref{hypersurg}. When  $\lambda >0$ (i.e. negative cosmological constant), 
the manifold $N$ is a circle and the Newton-Hooke spacetimes is called oscillating. When $\lambda =0$, 
 $N$ is the union of  two copies of  $\RR$  linearly embedded in $\RR^4$  and, picking one of these copies,  we recover the  Galilean spacetime reviewed in \cite{Papageorgiou:2009zc}.
For  $\lambda <0$  the manifold $N$ is   a  hyperbola (consisting of two branches) embedded in $\RR^4$  and the  Newton-Hooke spacetime is called expanding in this case. 

Newton-Hooke spacetimes have additional structures, inherited from the embedding in $\RR^4$
with the metric \eqref{met}. The manifold $N$ inherits the length element
\begin{equation}
\label{afftime}
d\tau^2 = \frac{1}{\lambda} dw^2 +  dt^2,
\end{equation} 
with an associated affine length parameter $\tau$ on $N$, which is defined up to an additive constant.  Note that, when $\lambda =0$, the defining equation in \eqref{hypersurg} is simply $w^2=1$. Hence $dw =0$ and the affine time $\tau$ is equal to the embedding time $t$,  up to an additive constant and a possible sign ambiguity.  For later use we note that the embedding coordinates $t$ and $w$ can be expressed in terms of the affine parameter $\tau$ in a unified fashion in terms of the formal parameter $\g{8}$,
\begin{align}
\label{embed}
t = \frac{\sin (\g{8} \tau)}{\g{8}}, \quad w = \cos(\g{8}\tau),
\end{align}
where we have chosen the arbitrary  additive constant such that $t=0$ when $\tau =0$.  Here $\sin$ and $\cos$ should be interpreted in terms of their power series; these are such that neither of the expressions in \eqref{embed} involves  odd powers of $\g{8}$.

Finally, the metric \eqref{met} also  induces a Euclidean line element
\begin{equation}
\label{euclid}
dr^2=dx^2 + dy^2
\end{equation}
on the spatial plane $\RR^2$, which does not depend  on $\lambda$: in the Newton-Hooke space time, space is flat, Euclidean space  regardless of the value of $\lambda$. 

We can recover the (unextended) Newton-Hooke groups defined in the previous section  as the sets of linear maps $\RR^4\rightarrow \RR^4$ which 
leave invariant the Newton-Hooke spacetimes \eqref{hypersurg} and preserve the spatial Euclidean structure \eqref{euclid}.  More precisely,  we concentrate on the component  in the set of all such maps  which also contains the identity. With the notation $\vec{x}= (x,y)^t$ and  $\vec{t}:=(\theta t,w)^t$ such maps necessarily have the form 
\begin{align}
\vec{t}& \mapsto R(\g{8}\alpha) \vec{t},  \nonumber \\
\vec{x}& \mapsto R(\varphi)\vec{x} +  t \vec{v} + w \vec{a},
\end{align}
with $\vec{v},\vec{a}\in \RR^2$ and $\varphi,\alpha \in \RR$ arbitrary parameters.  Here we used the abbreviation $R$ for the rotation matrix \eqref{rot}. Using the parametrisation \eqref{embed},
one can write this  map compactly in terms of the affine parameter $\tau$ as 
\begin{align}
\label{galtransf}
\tau&\mapsto \tau + \alpha,  \nonumber \\
\vec{x } &\mapsto R(\varphi) \vec{x} +  \frac{\vec{v}}{\g{8}}\sin (\g{8} \tau) +  \cos(\g{8}\tau) \vec{a}.
\end{align}
This action agrees with the action of the  Newton-Hooke group given  in the (2+1)-dimensional context in \cite{Alvarez:2007fw} and (for (3+1)-dimensional spacetimes) in
 \cite{NH_spacetimes_Gibbons}, where it is derived from the definition of Newton-Hooke spacetimes as cosets. In the next section,  we will also  use  the coset method in re-derive the law \eqref{galtransf}, exploiting the algebraic peculiarities of our (2+1)-dimensional situation. The derivation given above, starting from  linear maps in the ambient four-dimensional space appears to be new and gives a natural geometrical interpretation of  the formulae.  However, from   the embedding  point of view it is not immediately obvious that 
  that the composition law of such transformation reproduces the group law of the (unextended) Newton-Hooke group.  In the next section we will see that this follows trivially in  the alternative coset approach.

To end this section we comment on the geometry behind \eqref{galtransf}.  For $\lambda=0$  we recover  the usual Galilean transformation
 \begin{align}
\tau&\mapsto \tau + \alpha,  \nonumber \\
\vec{x } &\mapsto R(\varphi) \vec{x} +  \tau \vec{v} +   \vec{a}.
\end{align}
As the time parameter $\tau$ varies for fixed parameters $\varphi, \vec{v},\vec{a}$ of the Galilei transformation, the image of $\vec{x}$ traces out a straight line in  $\RR^2$ (space)  with uniform speed: these are the inertial motions in Galilean physics. The analogous inertial motions in the oscillating and expanding Newton-Hooke spacetimes can be read off from \eqref{galtransf}. They are, respectively,  ellipses and hyperbolae in  $\RR^2$. Further details, including the equations of inertial motion in Newton-Hooke spacetimes can be found  in \cite{NH_spacetimes_Gibbons}.

\subsection{Newton-Hooke spacetimes as cosets}

We now look at the classical spacetimes  as
symmetric spaces.
Recall  that  a symmetric space for  a Lie group $G$ is a homogeneous space $G/H$ where the stabiliser $H$ of a typical point is an open subgroup of the fixed point set of an involution (an automorphism which squares to the identity) of $G$. We now apply this  construction to the (unextended) Newton-Hooke group $\NHH$, using  the notation
\begin{equation}
  \label{eq:nwsptim}
  (\varphi,\vec{w}) \leftrightarrow \exp(\varphi J) \exp(\vec{w}
  \vec{J})
\end{equation}
 for elements of the (unextended) homogeneous Galilei group $\NWH$ and $ (\varphi + \g{8}\g{a},\vec{w} +  \g{8}\vec{a})$ for elements of  the Newton-Hooke group $\NHH$.
 The role of the involution is played by the `complex' conjugation
\begin{equation}
*: (\varphi + \g{8}\g{a},\vec{w} +  \g{8}\vec{a})\mapsto
 (\varphi - \g{8}\g{a},\vec{w} - \g{8}\vec{a}).
\end{equation} 
Clearly this is an involution of $\NHH$, and the fixed point set is
$\NWH$. We define the Newton-Hooke spacetimes as the homogeneous spaces 
\begin{equation}
M_\lambda= \NHH/\NWH.
\end{equation}
 To determine these, we need a factorisation of elements $g\in \NHH$ into elements $h\in \NWH$ and (unique) coset representatives $m\in \NHH$.

\begin{lemma}
Any element $g \in \NHH$ can be written uniquely as a product $g=m\cdot h$  of $h\in\NWH$ and an element of the form
\begin{equation}
\label{cosetpara}
m=(\g{8}\g{a}, \g{8} R(\g{8}\g{a}) \vec{a}) = (\g{8}\g{a}, \vec{0})
(0,\g{8}\vec{a}).
\end{equation}
\end{lemma}

\noindent {\bf Proof}: \,  Parametrising $h\in \NW$ as 
\[
h= (\varphi',\vec{w}')
\]
and $m$ as 
\[
m=(\g{8}\g{a}',\g{8} R(\g{8}\g{a}') \vec{a}')
\]
 we find that 
 \[
 (\g{ff} + \g{8}\g{a},\vec{w} +  \g{8}\vec{a})
 =m \cdot h
 \]
 iff
\begin{align}
  \label{eq:tilde-untilde}
  \g{a} =& \g{a}' & \vec{a} =& \cos({\g{8}\g{a}})\vec{a}' +
  \frac{\sin(\g{8}\g{a})}{\g{8}}\epsilon \vec{w'}\nonumber\\
\g{ff} =& \g{ff}' & \vec{w} =& \cos({\g{8}\g{a}})\vec{w}' + \g{l}\frac{\sin(\g{8}\g{a})}{\g{8}}\epsilon\vec{a}'.
\end{align}
Here  $\cos({\g{8}\g{a}})$ and $\sin(\g{8}\g{a})/ \g{8}$  are both real-valued functions of $\g{a}$ defined via their power series (which both only contain non-negative, even powers).  
The  linear relation between $\vec{w},\vec{a}$ and $\vec{w}',\vec{a}'$ has determinant=1, and is therefore invertible. It follows that every element $g\in \NHH$ can be factorised according to \eqref{cosetpara} in a unique fashion. 
\hfill $\Box$

The factorisation allows us to define an action of the Newton-Hooke
groups on the Newton-Hooke space times as follows. We introduce a simplified notation for elements of   $\NHH$  
\[
(\g{ff},\vec{w};\g{a},\vec{a}):=(\g{8}\g{a},\g{8} R(\g{8}\g{a}) \vec{a})(\varphi,\vec{w})
\]
and space-time points
\[
[\tau, \vec{x}]:=(\g{8}\tau, \g{8}R(\g{8}\tau)\vec{x}).
\]
We then define the action $\rho$ of $\NWH$  via
\begin{equation}
\rho ((\g{ff},\vec{w};\g{a},\vec{a})): [\tau, \vec{x}]
\mapsto [\tau', \vec{x}'],
\end{equation}
where  $ [\tau', \vec{x}']$ is defined via the factorisation
\[
(\g{8}\g{a}, \g{8}R(\g{8}\g{a}) \vec{a})(\varphi,\vec{w})(\g{8}\tau, \g{8}R(\g{8}\tau)\vec{x})
=(\g{8}\tau', \g{8} R(\g{8}\tau')\vec{x}')(\varphi',\vec{w}').
\]
Computing the products and the factorisation one finds that 
\begin{equation}
  \label{spacetime_symmetries}
 \rho((\g{ff},\vec{w};\g{a},\vec{a})): [\g{t}, \vec{x}] \mapsto
[\g{t} +\g{a}, R(\varphi)\vec{x} + \dfrac{\vec{v}}{\g{8}} \sin
(\g{8}\g{t}) + \vec{a} \cos (\g{8}\g{t})],
\end{equation}
in precise agreement with \eqref{galtransf}.  In the current formulation in terms of factorisations it is manifest from the start that \eqref{spacetime_symmetries} defines a group action of the Newton-Hooke group, thus filling the gap in the argument at the end of the previous section.

\section{The Chern-Simons action for Galilean gravity with cosmological constant}
In this short section we  introduce an action for Galilean gravity with cosmological constant in 2+1 dimensions. We follow the approach of \cite{Papageorgiou:2009zc} and refer the reader to that paper for details. We start with a spacetime manifold $\manif$ and our goal
 is to construct a dynamical model on $\manif$ whose solutions equip open subsets of $\manif$ with the structure of a Newton-Hooke space time which are glued together  using Newton-Hooke transformations. We achieve this by using a Chern-Simons action for the doubly-extended Newton-Hooke group. Solutions are flat connections $A$ which can be trivialised, in open neighbourhoods, in  terms of maps $g: U\subset \manif \rightarrow \NH$  i.e., locally we have $A=g^{-1}\dd g$.  The translation parts of these maps provide the local identification of $U$ with a portion of a Newton-Hooke spacetime.

Specifically, the  gauge field of the Chern-Simons action is locally a one-form on
spacetime $\manif$ with values in the Lie algebra of
the extended Newton-Hooke group. We write it as 
 \begin{equation}
 \label{gaugefield}
 A = \omega  J +\epsilon_{ij} \omega_j J_i  + \eta S + e H + e_i P_i + f M,
 \end{equation}
 where $\omega, \omega_1, \omega_2, \eta, e,  e_1, e_2$  and $f$ are
 ordinary one-forms on $\manif$ and we have chosen our notation to agree with that used in \cite{Papageorgiou:2009zc}.
 Then one checks straightforwardly that 
 the  curvature  can be written as a sum
 \begin{equation}
\begin{split}
 \label{curvsplit}
 F=&\dd A +\frac 1 2 [A,A]  = R + C + T,\\
\end{split}
 \end{equation}
 where the homogeneous term $ R $ and the cosmological term $ C $ take values in $\nawi$,
 \begin{equation}
\begin{split}
R=& \dd\omega \; J+ (\dd\eta + \omega_1\wedge \omega_2) S +(\epsilon_{ji}\dd\omega_i 
+\omega_{j}\wedge \omega) J_j\\
C=& \g{l} e_i\wedge e\g{e}_{ij} J_j +\g{l}e_1\wedge e_2S
\end{split}
\end{equation}
while the  torsion term  is
\begin{equation}
T = \dd e\; H  + (\dd f +\omega_i\wedge e_i) M +(\dd e_j +\omega_j\wedge e + \epsilon_{ij}e_i\wedge \omega)P_j.
\end{equation}

 In order to write down a Chern-Simons action for the gauge field \eqref{gaugefield} we require an invariant, non-degenerate bilinear form on $\nawi$.  We shall use the pairing
\begin{align}
  \label{eq:inprod_nh}
  \lie{J}{M}=\lie{H}{S}=\frac{1}{8 \pi G}, \quad  \lie{P_i}{J_j}=-\g{d}_{ij}\frac{1}{8 \pi G},
\end{align}
which  depends on Newton's constant $G$ (which has physical dimension of  inverse mass)  but is independent of $\g{l}$. It  was already  discussed and used in  \cite{Papageorgiou:2009zc} where, however, we did not absorb Newton's constant in it. We are doing this in the current paper since it allows for a more transparent discussion of how the various physical parameters enter the quantisation. With $G$ included as above, the dimension of the pairing $\lie{\cdot}{\cdot}$ is that of an action.   Our Chern-Simons action for Galilean  gravity  with cosmological constant in
 2+1 dimensions is
\begin{align}
\label{CSact}
I_{CS}[A] = &\frac 1 2  \int_\manif \langle A \wedge \dd A\rangle +\frac 1 3 \langle A \wedge [A,A]\rangle 
 \nonumber \\
=&\frac 1 2 \biggl( \int_\manif \omega \wedge \dd f + f\wedge
   \dd\omega + e\wedge \dd\eta + \eta\wedge \dd e +\epsilon_{ij}(
   \omega_i\wedge \dd e_j + e _j\wedge \dd \omega_i) \nonumber \\
   &+  2 \int_\manif \omega \wedge 
      \omega_i \wedge e_i+ e\wedge \omega_1\wedge \omega_2 + \g{l} e \wedge e_1\wedge e_2  \biggr).
\end{align}
Note in particular that the cosmological term is proportional to the space-time volume $\int_\manif e\wedge e_1\wedge e_2$, as expected.  Also, we recover the action for Galilean gravity without cosmological constant discussed in \cite{Papageorgiou:2009zc} as $\lambda\rightarrow 0$.

Restricting attention to the case where $\manif$ is a product of a two-dimensional manifold $\Sigma$ (representing space) and $\RR$ (representing time), the 
the phase space of Chern-Simons theory on  $\manif$ is the moduli  space of flat connections on $\Sigma$, equipped with the Atiyah-Bott symplectic structure \cite{AB,Atiyah}.  One can parametrise the phase space in terms of holonomies around all non-contractible paths on $\Sigma$ relative to a fixed base point, modulo conjugation. Punctures on the spatial manifold $\Sigma$ represent particles and need to be decorated with co-adjoint orbits of the gauge groups (here  the extended Newton-Hooke group).  These orbits are labelled by the masses and spins of the particles. Further details about the interpretation of the holonomies  in (2+1)-dimensional gravity as well as  further references  can be found in \cite{Schroers, MS1,schroers-2007-035}.  For our purposes it is important that in a scheme invented by Fock and Rosly  \cite{FR} a Poisson bracket on the extended phase space  of all holonomies  is given in terms of a classical $r$-matrix which solves the classical Yang-Baxter equation and which is compatible with the  inner product used in the Chern-Simons action.  The compatibility requirement is that  symmetric part of the  $r$-matrix equals the Casimir element associated with the inner product used in defining the Chern-Simons action. Fock and Rosly's description of the phase space of Chern-Simons theory is the starting point of the combinatorial or Hamiltonian quantisation programme \cite{AGSI,AGSII,AS} where the Hilbert  space of the quantised theory is constructed in terms of the representation theory of a  quantum group whose quantum $R$-matrix reduces to the classical $r$-matrix of  Fock and Rosly in the classical limit. In the context of (2+1)-dimensional gravity, this quantum group deforms the local isometry group of the classical theory \cite{Schroers, BNR, MS2}.

\section{The extended Newton-Hooke algebra as a classical double}

In this section we  exhibit a bi-algebra structure on the  doubly extended Newton-Hooke Lie
algebra  which is compatible with the Chern-Simons action of the previous section in the sense of Fock and Rosly.   There is now a unified picture of bi-algebra structures on symmetry algebras of 2+1 dimensional spacetimes, parametrised by the speed of light and the cosmological constant \cite{BHdOS_q,
Ballesteros2010375}. However, in the current context we are interested in the centrally extended  Newton-Hooke algebra, which was not considered in \cite{BHdOS_q,
Ballesteros2010375}. 

\subsection{The extended Heisenberg bi-algebra}
The Lie algebra $ \nawi $ admits a one-parameter family of non-degenerate, ad-invariant  bilinear
forms $ \lie{}{}_{\nawi} $ of Lorentzian signature
\begin{align}
  \label{inprod_g0}
  \lie{J}{S}_{\nawi}=&\frac 1 \aa & \lie{J_i}{J_j}_{\nawi}=&-\g{d}_{ij}\frac 1 \aa,
\end{align}
which has played a role in its applications in string theory \cite{Nappi:1993ie,FoFS}.
The  arbitrary, possibly complex, parameter $\aa \neq 0$  is important for us and will be fixed later. 
 In the Heisenberg basis \eqref{eq:nawi_z} we have 
\begin{align}
\label{inprodnew}
\lie{\newS}{\newJ}_{\nawi} = &  \frac 2 \aa , & \lie{Z}{\bar{Z}}_{\nawi} =& -\frac  2 \aa.
\end{align}

The pairing \eqref{inprod_g0} is not  used  directly in this paper  but various structures that we encounter are closely related to it. At this point we note that
 that the pairing \eqref{eq:inprod_nh} on the extended Newton-Hooke Lie algebra, which is central to our discussion, can be obtained, for a suitable value of $a$,  by   extending \eqref{inprod_g0}  $ \mathcal{R}_{\g{l}} $-linearly  and then taking the terms linear in $\theta$ (the `imaginary part').

For the remainder of this and the next section we work with the basis $\{\newH,Z,\bar Z,\newS\}$  of $\heis$. The  Casimir associated to \eqref{inprodnew}  is
\begin{equation}
K = \frac{\aa}{2}\left(\newS\otimes \newJ + \newJ\otimes \newS\right) - \dfrac{\aa}{2} \left(\bar{Z} \otimes Z +Z \otimes \bar{Z}\right).
\end{equation}
The cubic Casimir
\begin{equation}
\Omega = \bar{Z}\wedge Z \wedge \newS
\end{equation}
is related to the quadratic Casimir via the classical Yang-Baxter relation
\begin{equation}
[[K,K]] = \frac {\aa^2} {4 } \Omega.
\end{equation}
An easy calculation shows that 
\begin{equation}
r_A=  \frac{\aa}{2}\bar{Z}\wedge Z
\end{equation}
satisfies
\begin{equation}
[[r_A,r_A]] = -\frac { \aa^2 }{ 4}\Omega.
\end{equation}\\
It follows from general results (see e.g.   page 54 in  \cite{CP})
that any scalar multiple of  $K \pm  r_A$ satisfies the classical Yang-Baxter equation
$ [[r,r]]=0$. For us, the combination 
\begin{equation}
\label{eq:rmatrix}
r= -a(K +r_A) = -\frac{\aa}{2}\left(\newS\otimes \newJ +\newJ\otimes \newS\right) +\aa Z\otimes  \bar{Z} 
\end{equation}
will be important. 
Computing  the co-commutators via 
\begin{equation}
\delta(X)=(1\otimes\ad_X + \ad_X\otimes 1)(r)
\end{equation}
one finds  that they only depend on the antisymmetric part  $r_A$  and are given by 
\begin{equation}
\label{cobracks}
\delta(Z) = -\frac{\aa}{2} \newS\wedge Z \quad \delta(\bar{Z}) = -\frac{\aa}{2}\newS\wedge\bar{Z},
\end{equation}
with all others being zero. The 
induced dual  Lie algebra structure $\heis^*$   has  the brackets
\begin{align}
\label{dualbrackets}
 [\newS^*,Z^*]^*&=- \frac{\aa}{2} Z^*, \quad [\newS^*,\bar{Z}^*]^*=- \frac{\aa}{2}
 \bar{Z}^*,  \nonumber \\
  [Z^*,\bar{Z}^*]^*&= [\newJ^*,Z^*]^* = [\newJ^*,\bar{Z}^*]^*=[\newJ^*,\newS^*]^*=0.
\end{align}

\subsection{The extended Newton-Hooke Lie algebra as a classical double of the extended Heisenberg algebra}
\label{newtonhookebi}

We now show that one can find a subalgebra of the complexification $\nh\otimes \CC$ which is isomorphic to the Lie algebra $\heis^*$ and complements the 
subalgebra $\heis$ of $\nh \otimes\CC$  in the sense that  
   $ \nh \otimes\CC= (\heis \oplus \heis^*)\otimes \CC  $. Moreover we will see that, with respect to the pairing
 \eqref{eq:inprod_nh}, the subalgebras  $\heis$  and $\heis^*$ are each null  but in duality to each other. Together, these results amount to showing that $\nh\otimes \CC$ has the bi-algebra structure of a classical double of  the bi-algebra $\heis$ (with the bi-algebra structure of the previous section).

 Introducing the (generally complex) linear combinations
\begin{align}
\label{basisch}
\Pi_i =& P_i + \sqrt{-\lambda} \epsilon_{ij} J_j & \newM =& -2\ii M &
\newH = & \ii H
\end{align} 
and defining 
\begin{equation}
\g{P} = \g{P}_1 + \ii \g{P}_2, \quad \bgp = \g{P}_1 - \ii \g{P}_2,
\end{equation}
we note that the generators $H,M,\Pi,\bgp$ span a
Lie subalgebra of   $\nh\otimes \CC$ with brackets 
\begin{align}
\label{dualinside}
 [\newH,\Pi] &= -\sqrt{\lambda} \Pi, \quad
[\newH,\bgp] = -\sqrt{\lambda} \bgp, \nonumber \\ [\Pi,\bgp] &= [\newM,\Pi]=[\newM, \bgp]=[\newM,\newH]=0.  
\end{align}
Up to a scale, which we will fix presently, this agrees with the brackets of \eqref{dualbrackets} of  the Lie algebra $\heis^*$. It is a trivial
central extension  of Lie algebra generated by $\newH,\Pi_1$ and
$\Pi_2$. Thus the  Lie algebra $\nh\otimes \CC$  has, in addition to the subalgebra $\heis$,
a Lie  subalgebra which is isomorphic to $ \heis^*$ and which will also denote by $\heis^*$. As a vector space,  $ \nh \otimes\CC= (\heis \oplus \heis^*)\otimes \CC  $ but the brackets between generators of different subalgebras are not trivial, with non-zero brackets given by
\begin{align}
\label{eq:com_relations_between_algebras}
 [\bar Z,\g{P}]= &- \newM-\sqrt{\lambda} \newS,
& [Z , \newH] = &\g{P} - \sqrt{\lambda} Z,
& [\g{P},\newJ]= &\g{P}, \nonumber \\
[ Z,\bgp]= & \newM- \sqrt{\lambda} \newS ,
& [\bar Z , \newH] = & -\bgp -\sqrt{\lambda} \bar Z,
& [\bgp,\newJ]= &- \bgp.
\end{align}
The quickest way to check these relations is to note that the generators $ \g{P}, \bgp $  can be
written in terms of the parameter $\theta$  as
\begin{align}\label{eq:Pi_Z_relations}
  \g{P} =& (\g{8} + \sqrt{\g{l}})Z &\bgp =& (\g{8} - \sqrt{\g{l}})\bar{Z}.
\end{align}

The pairing in the new basis can be obtained from \eqref{eq:inprod_nh}, with  the non-zero pairings given by
\begin{align}
\label{eq:inprod_Z}
 \lie{Z}{\bgp} =& \lie{\bar{Z}}{\g{P}} = -\frac{1}{4\pi G},& \lie{\newM}{\newJ} = \lie{\newS}{\newH} = \frac{1}{4\pi G}. 
\end{align}
We see that the Lie subalgebras  $\heis$ and $\heis^*$   are both null  with respect to $\lie{\cdot}{\cdot}$ (the pairing of any of its generators with any other of its generators is zero) and   dual to each other. In particular, the pairing 
 \eqref{eq:inprod_Z} gives the following  identification of the dual basis to 
the $\heis$-basis $\{\newJ,\newS, Z,\bar{Z}\}$: 
\begin{equation}
\begin{aligned}
\label{basisid}
\newJ^* &\simeq 4\pi G \newM, &&& \newS^* &\simeq 4\pi G \newH, &&& Z^* &\simeq -4\pi G \Pi^{\circ}, &&& \bar{Z} &\simeq - 4\pi G \Pi,
\end{aligned}
\end{equation}
 Comparing the brackets \eqref{dualinside} and \eqref{dualbrackets} we  see that, in order to obtain agreement with the identifications  \eqref{basisid}, we need to fix the scale parameter $\aa$ to be
 \begin{equation}
 \label{firstpara}
 \aa = 8\pi \sqrt{\lambda} G.
 \end{equation}

We have thus shown that $\nh\otimes \CC$ has the Lie algebra structure of a classical double of $\heis$.  The canonical bi-algebra structure on this double is  determined by the canonical
 $r$-matrix \cite{CP} which takes the form
\begin{equation}
  \begin{split}
\label{doubler}
  r_{\nh} =& 4\pi G \left(\newM \otimes \newJ + \newH\otimes \newS - \bar \Pi \otimes Z + \Pi
  \otimes \bar Z\right) \\
  =&  8\pi G \left(M \otimes J+H\otimes S - \Pi_i \otimes J_i\right)  \\
  =&  8\pi G \left(M \otimes J+H\otimes S - P_i \otimes J_i
  -\sqrt{-\lambda}J_1\wedge J_2\right)
  \end{split}
\end{equation}
It follows from the general theory of classical doubles \cite{CP} that $r_{\nh}$
satisfies the classical Yang-Baxter equation.  One can also check this using the parameter $\theta$ by noting that
\[ r_{\nh} =-\frac {1} {2 \sqrt{\lambda}}  (\theta\otimes 1 + \sqrt{\lambda}\, 1 \otimes 1) r - \frac {1} {2\sqrt{\lambda}}  (\theta\otimes 1 - \sqrt{\lambda}\,1\otimes 1) r', 
\]
and using that $r$ and the flipped $r$-matrix $r'$ satisfy the classical Yang-Baxter equation and that 
\[
(\theta\otimes 1 + \sqrt{\lambda}\, 1 \otimes 1) (\theta\otimes 1 - \sqrt{\lambda}1\otimes 1)=0.
\]

Thus we have found a classical $r$-matrix for the Lie algebra $\nh$ which whose symmetric part equals the Casimir associate to the pairing \eqref{eq:inprod_nh} and which is thus compatible with the Chern-Simons action \eqref{CSact} in the sense of Fock and Rosly. Moreover, the $r$-matrix  $r_{\nh}$ is that of a classical double. 

In this section we have worked with the complexification  $\nh\otimes \CC$ and displayed its structure as a double of the $\heis$. This point of view turns out to be the most convenient for the quantisation which we study in the next section. However, it is worth noting that  when $\lambda <0$  the (uncomplexified)  bi-algebra $\nh$ can be viewed as a classical double of the (uncomplexified) algebra $\nawi$, with the basis of the $\nawi^*$ provided by $\{M,H,\Pi_1,\Pi_2\}$.  This is clear from the reality of the basis change $P_i\rightarrow \Pi_i$ \eqref{basisch} when $\lambda <0$.

\section{Quantisation of the extended Heisenberg bi-algebra}

\subsection{The extended $q$-Heisenberg
  algebra and its role in Galilean gravity}

We now turn to the 
quantisation of  the algebra $\heis $ \eqref{eq:nawi_z}  with the bi-algebra structure determined by the the classical $ r$-matrix \eqref{eq:rmatrix}. Happily, this problem is much studied, and it is known that a quantisation is provided by a certain $q$-deformation of the extended Heisenberg algebra. The  $q$-deformed Heisenberg algebra and its extension was first defined and studied in \cite{CGST} and further investigated   in \cite{GS}.
A later, systematic study of different bi-algebra structures  on the extended Heisenberg or  oscillator algebra and their quantisation can be found in \cite{BH}. Here
we adopt the approach and notation of the  textbook  treatment in \cite{Majid:1996kd}.   The main purpose of this section is to translate the results in  \cite{Majid:1996kd} into our language, and, crucially, to connect the parameter $q$ to the physical parameters entering Galilean quantum gravity with a cosmological constant. 

In \cite{Majid:1996kd} the generators of the extended Heisenberg algebra $\heis$ are denoted $ \{a, a^\dagger, N, H\}$; they are related to our generators $\{\bar Z,Z, \newJ, \newS\}$ via  the identification
\begin{align}
  \label{eq:id_with _majid}
  a \leftrightarrow & Z & a^\dagger \leftrightarrow & \bar{Z} & N \leftrightarrow& \newJ & H \leftrightarrow& \newS. 
\end{align}
For the convenience of the reader, we reproduce  the definition of  Example 3.1.2,
 from  \cite{Majid:1996kd} here in our notation:

Let $ q $ be a nonzero parameter. The $ q$-Heisenberg algebra
is  defined with generators $ Z, \bar{Z}, q^{\frac{\newS}{2}},
q^{-\frac{\newS}{2}} $ and $ 1 $ with the relations $
q^{\pm\frac{\newS}{2}}q^{\mp\frac{\newS}{2}} = 1 $ and
\[ [q^{\frac{\newS}{2}}, Z] = 0, \qquad [q^{\frac{\newS}{2}}, \bar{Z}] = 0,
\qquad [Z, \bar{Z}] = \frac{q^{\newS} - q^{-\newS}}{q - q^{-1}}.\]
This forms a Hopf algebra with co-product
\[\g{D}Z = Z\otimes q^{\frac{\newS}{2}} + q^{-\frac{\newS}{2}}\otimes Z,
\qquad \g{D}\bar{Z} = \bar{Z}\otimes q^{\frac{\newS}{2}} + q^{-\frac{\newS}{2}}\otimes \bar{Z}\]
\[\g{D}q^{\pm\frac{\newS}{2}} = q^{\pm\frac{\newS}{2}}\otimes
q^{\pm\frac{\newS}{2}}, \qquad \g{e}q^{\pm\frac{\newS}{2}} = 1, \qquad \g{e}Z
= 0 = \g{e} \bar{Z} \]
and antipode
\[\mathcal{S} Z = -Z, \quad \mathcal{S} \bar{Z} = - \bar{Z}, \quad \mathcal{S}q^{\pm\frac{\newS}{2}} =
q^{\mp\frac{\newS}{2}}.\]
The extended $ q$-Heisenberg algebra
is defined with the additional mutually inverse generators $ q^{\newJ},
q^{-\newJ} $ and relations
\[  q^{\newJ}Z
q^{-\newJ} = q^{-1}Z, \qquad q^{\newJ}\bar{Z}q^{-\newJ} = q\bar{Z}, \qquad [q^{\newJ},
q^{\frac{\newS}{2}}] = 0.\]
It forms a Hopf algebra with the additional structure
\[\g{D}q^{\newJ} =
q^{\newJ}\otimes q^{\newJ}, \qquad \g{e}q^{\newJ} = 1, \qquad \mathcal{S}q^{\pm \newJ}=q^{\mp \newJ}.\]
If $ q = e^{\frac{t}{2}} $ and we work over $ \CC[[t]] $ rather than $
\CC $ then we can regard $ Z, \bar{Z}, \newS, \newJ $ and $ 1 $ as the
generators. In this case, the extended $  q$-Heisenberg Hopf algebra
is quasitriangular with
\[\mathcal{R} = q^{-(\newJ\otimes \newS + \newS\otimes \newJ)}e^{(q -
  q^{-1})q^{\frac{\newS}{2}}Z\otimes q^{-\frac{\newS}{2}}\bar{Z}}\]


In order to show that this Hopf algebra quantises the bi-algebra $\heis$ with the $r$-matrix 
\eqref{eq:rmatrix}  we first need to look carefully at the physical dimensions of  the generators.  Recall that both $Z$ and $\bar Z$ have dimension of inverse velocity and that $ \newS $ has the dimension of inverse velocity squared. In order to make sense of the exponential of $\newS$ we therefore need a parameter of dimension velocity squared, which we call $\ha$. Then the combination $\ha \newS$ is dimensionless. We can then regard the $q$-deformed Heisenberg algebra as generated by $\{Z,\bar Z, e^{\frac{\ha}{2} \newS},
e^{-\frac{\ha}{2} \newS}, \newJ, 1\}$  and with the  algebra
\eqref{eq:nawi_z} replaced by  the relations
\begin{align}
  \label{eq:q_heis}
  \comm{Z}{\bar{Z}} =& \frac{e^{\frac{\ha}{2} \newS} - e^{-\frac{\ha}{2} \newS}}{\ha} & 
  \comm{Z}{\newJ} =& Z & \comm{\bar{Z}}{\newJ} =& -\bar{Z},
\end{align}
as well as $e^{\pm\frac{\ha}{2}\newS}e^{\mp\frac{\ha}{2}\newS} = 1 $.
The co-algebra can be written as 
\begin{equation}
\begin{aligned}
\label{eq:cocom_qNW}
  \g{D}(Z) & = Z\otimes e^{ \frac{\ha}{4} \newS} + e^{ -\frac{\ha}{4} \newS}\otimes Z &
  \g{D}(\bar{Z}) & = \bar{Z}\otimes e^{ \frac{\ha}{4} \newS} + e^{- \frac{\ha}{4} \newS}\otimes \bar{Z}\\ 
\g{D}(e^{ \pm\frac{\ha}{4} \newS}) & = e^{ \pm \frac{\ha}{4} \newS} \otimes e^{ \pm
  \frac{\ha}{4} S} & \g{D}(\newJ) & = 1\otimes \newJ + \newJ \otimes 1
\end{aligned}
\end{equation}
while the antipode $ \mathcal{S} $ and counit $ \g{e} $ are
\begin{equation}
\begin{aligned}
\label{eq:ant_qNW}
  \mathcal{S} (Z) & =  -Z&
  \mathcal{S} (\bar{Z}) & = -\bar{Z}\\ 
\mathcal{S} (e^{ \pm \frac{\ha}{4} \newS}) & = e^{ \mp \frac{\ha}{4} \newS} & \mathcal{S} (J)
& = {- J}\\
 \g{e}(Z) = \g{e}(\bar{Z})  & =0  &
\g{e}(e^{ \mp \frac{\ha}{4} \newS}) = \g{e}(e^{ \mp \frac{\ha}{4} \newJ}) & = 1,
\end{aligned}
\end{equation}
and  the quantum $ \mathcal{R}$-matrix is
\begin{equation}
  \label{eq:quantum_r}
  \mathcal{R} = \exp\bigl(- \frac{1}{2}\ha (\newJ\otimes \newS + \newS \otimes
  \newJ)\bigr)\exp\bigl(\ha e^{\frac{\ha}{4} \newS}Z\otimes e^{ -\frac{\ha}{4} \newS}\bar{Z}\bigr).
\end{equation}
We use the notation $U_\ha(\heis)$ for the $q$-deformed Heisenberg algebra in this notation. 
Expanding the quantum R-matrix in the parameter $\ha$  and keeping at most linear terms we have
\begin{equation}
 \mathcal{R} \simeq 1\otimes 1 -\frac{\ha}{2}\left(\newS\otimes \newJ +\newJ\otimes \newS\right) + \ha Z\otimes  \bar{Z}.
\end{equation}
This is indeed of the form
\begin{equation}
\label{eq:r_expand}
\mathcal{R} \simeq 1\otimes 1 + \hbar r,
\end{equation}
where $r$ is the $r$-matrix \eqref{eq:rmatrix}, provided we set
\begin{equation}
\label{paraid}
\ha = \hbar \aa =  8\pi \hbar G \sqrt{\lambda},
\end{equation}
where we used \eqref{firstpara}. Thus we have shown
that the $U_\ha (\heis)$ quantises the bi-algebra $\heis$ with classical $r$-matrix \eqref{eq:rmatrix} and, at the same time,  deduced  the  important identification  \eqref{paraid} of the deformation parameter $\ha$ with the physical constants $\hbar,G,\lambda$ which enter Galilean quantum gravity with a cosmological constant.   One checks that $\ha$ has the required dimensions of velocity squared.\\

\subsection{Real structures}

In our treatment thus far we have permitted ourselves to  consider Lie algebras over the complex numbers when convenient. Thus, although for our applications the  basis  of $\nh$ used in \eqref{NewtonHookeimproved}  has the clearest physical interpretation, we have mostly worked  with the Heisenberg generators and their duals  when discussing the bi-algebra structure of $\nh$ in Sect.~\ref{newtonhookebi}. We now need to select the appropriate real form with the help of $*$-structures.   Recall that a $*$-structure on a complex Hopf algebra $A$ is an anti-linear anti-algebra-automorphism $A\rightarrow A$ which  squares to the identity and which is compatible with the co-product in the sense that $\Delta(h^*) = (\Delta h)^{*\otimes *}$ for any $h\in A$. A full list of axioms can be found in \cite{Majid:1996kd}.  In representations of $A$  on a Hilbert space  one  demands that a  $*$-structure be represented by taking the adjoint. Thus, in applications to quantum mechanics we need a $*$-structure to characterise Hermitian operators, i.e.  observables. 
For the extended $q$-Heisenberg algebra possible $*$-structures are listed in \cite{Majid:1996kd}. We need to distinguish two cases. 

First consider the case $ \g{l} > 0 $  so that  $ \ha $ is real. In this case there are two possible $*$-structures on $U_\ha(\heis)$.  In order to pick  one we recall the requirements coming from the physics. For unitary representations of the extended homogeneous Galilei algebra $\nawi$  we would expect the spin and angular momentum generators $S$ and $J$ to be represented by anti-Hermitian operators, so that $\newS $ and $\newJ$ are Hermitian. The boost generators $J_1$ and $J_2$ should both be represented by either Hermitian or anti-Hermitian operators (we can switch from one to other by multiplying with $i$ and absorbing the extra minus sign in the commutator $[J_1,J_2]$ in a re-definition of $S$). We thus expect $Z$ and $\bar Z$ to be mapped into each other under the $*$ operation. 
For $\lambda >0$ this can indeed be achieved by working with  the  $ *$-structure
\begin{equation}
  \label{eq:star_q_NW}
  Z^* = \bar{Z} \quad \bar{Z}^* = Z, \quad \newJ^* = \newJ, \quad \newS^* =  \newS,
  \quad (e^{ \frac{\ha}{2} \newS})^* = e^{ \frac{\ha}{2} \newS}.
\end{equation}
The $ R$-matrix is real in the sense that 
\[ \begin{split}
\mathcal{R}^{*\otimes*} =& \exp\bigl((e^{
    \frac{\ha}{2}} -  e^{-\frac{\ha}{2}})^* (e^{\frac{\ha}{2} \newS}Z\otimes e^{ -\frac{\ha}{2}
    \newS}\bar{Z})^{*\otimes *}\bigr)\exp\bigl( \ha (\newJ\otimes \newS + \newS \otimes
  \newJ)\bigr)\\
=& \exp\bigl((e^{
    \frac{\ha}{2}} -  e^{-\frac{\ha}{2}}) \bar{Z}e^{ \frac{\ha}{2} \newS}\otimes Z e^{ -\frac{\ha}{2}
    \newS}\bigr)\exp\bigl( \ha (\newJ\otimes \newS + \newS \otimes
  \newJ)\bigr)\\
=& \g{s}(\mathcal{R})
\end{split}
\]
where $  \g{s} $ is the flip operator.

Next we consider the the case $ \g{l} < 0 $ i.e. $ \ha \in \ii\RR$. In this case there appears to be a unique choice of $*$-structure on  $U_\ha(\heis)$  \cite{Majid:1996kd},  which is given by 
\begin{equation}
  \label{eq:anitr_star_q_NW}
  Z^* = Z \quad \bar{Z}^* = \bar{Z}, \quad \newJ^* =  - \newJ, \quad \newS^* = - \newS,
  \quad  (e^{ \frac{\ha}{2} \newS})^* = e^{ \frac{\ha}{2} \newS}.
\end{equation} 
With this structure one finds that $\mathcal{R}^{*\otimes *} = \mathcal{R}^{-1}$
so that, in the terminology of  \cite{Majid:1996kd}, the algebra is antireal quasitriangular. 

We argued above that, based on their role in the  Galilei algebra, we expect $Z$ and $\bar Z$ to be mapped into each other by a $*$ operation. The $*$-structure \eqref{eq:anitr_star_q_NW} does not do this and  is difficult to make sense of since  it amounts to $(J_1)^*=J_1$ and $(J_2)^*=-J_2$ i.e. to a different treatment of the two boost operators. The ultimate role of this $*$-structure in Galilean quantum gravity with $\lambda <0$ therefore remains unclear.

To end our discussion of $*$-structures we should alert the reader to a potential source of confusion when comparing our discussion with that in the literature on relativistic quantum gravity in 2+1 dimensions. The standard result there is that $D(U_q(sl_2(\RR))$, with $q$ on the unit circle,  controls 2+1 quantum gravity with a negative cosmological constant $\Lambda$ (corresponding to our $\lambda >0$) and that $D(U_q(su(1,1))$, with $q\in \RR$, 
 controls (2+1)-dimensional quantum gravity with a positive cosmological constant $\Lambda$  (corresponding to our $\lambda <0$). The deformation parameter $q$ in those discussions is
\[
 q= e^{- \frac{\hbar G \sqrt{c^2\Lambda}}{c^2}},
\]
where $c$ is the speed of light, and  Euclidean gravity can be included in discussion by taking $c^2 <0$.  It is clear that after taking the Galilean limit no constant of dimension  velocity is available and  that one cannot form a dimensionless quantity out of $\hbar, G$ and $\lambda$.  Instead, $q$-deformation in the Galilean limit involves exponentiating the generator $\newS$, which has dimension of inverse velocity squared. Recalling that $\newS = -2\ii S$ and  that $S$ is the generator obtained from a trivial central extension of the Lorentz group \cite{Papageorgiou:2009zc},  we  note that  the combination 
\[
e^{\frac{\ha}{2}\newS} = e^{-8\pi i\hbar G\sqrt{\lambda} S} = e^{-8\pi \hbar G\sqrt{-\lambda} S}
\]
appearing in \eqref{eq:q_heis} can formally be written as a $q$-exponential $(\tilde q)^S$, with $\tilde q$
indeed real when $\lambda <0$ and on the unit circle when $\lambda >0$, just like in the relativistic situation.

\section{Outlook and discussion: the quantum double $ D(U_\ha (\heis))$}

Since the the extended Newton-Hooke bi-algebra structure is that of  a classical double of a certain bi-algebra structure on $\heis$ one expects, on general grounds \cite{Semenov}, its quantisation to be given by the quantum double  $D(U_\ha (\heis))$ of the extended $q$-Heisenberg algebra of the previous section. This quantum double, which we will refer to as the Newton-Hooke double, does not appear to have been studied in the literature. The case $\ha=0$ was considered in \cite{Papageorgiou:2009zc}, where it was called the Galilei double. However, the inclusion of $\ha\neq 0$ presents a technical challenge. 

As a vector space,  the
  quantum double $D(U_\ha (\heis))$  has  the  structure $ U_\ha (\heis) \otimes (U_\ha (\heis))^*$. The dual quantum
  group $ (U_\ha (\heis))^* $ has itself been studied 
  extensively \cite{HLR} and is called the $q$-Heisenberg group.  It can be viewed as a deformation of the algebra of algebraic functions on the extended Heisenberg group  $\HEIS$ in the matrix representation \eqref{eq:matrixnawi}.  As shown in \cite{HLR}, the $q$-Heisenberg group can  be  constructed by the $ R$-matrix method \cite{Majid:1996kd}. It is generated by four real generators  $ A, B, C, D $,  which should be thought of as the matrix entries of 
\[ T =\left(
\begin{array}{lll}
 1 & A & B \\
 0 & C & D \\
 0 & 0 & 1
\end{array}
\right).\]
Computing the $ R$-matrix  by expressing \eqref{eq:quantum_r} in  the 
  representation \eqref{eq:matrices}
\begin{align}
  \label{eq:2}
  R = \left(
\begin{array}{ccccccccc}
 1 & 0 & 0 & 0 & 0 & 0 & 0 & 0 & 0 \\
 0 & 1 & 0 & 0 & 0 & \ha  & 0 & -\frac{\ha}{2}  & 0 \\
 0 & 0 & 1 & 0 & 0 & 0 & 0 & 0 & 0 \\
 0 & 0 & 0 & 1 & 0 & -\frac{\ha}{2}  & 0 & 0 & 0 \\
 0 & 0 & 0 & 0 & 1 & 0 & 0 & 0 & 0 \\
 0 & 0 & 0 & 0 & 0 & 1 & 0 & 0 & 0 \\
 0 & 0 & 0 & 0 & 0 & 0 & 1 & 0 & 0 \\
 0 & 0 & 0 & 0 & 0 & 0 & 0 & 1 & 0 \\
 0 & 0 & 0 & 0 & 0 & 0 & 0 & 0 & 1
\end{array}
\right),
\end{align}
the relations between the generators $A,B,C$ and $D$ are determined by the matrix equation
\[R T_1 T_2 = T_2 T_1 R,\]
with $T_1=T\otimes 1$ and $T_2=1\otimes T$. 
Explicitly, one finds
\begin{align}
  \label{eq:commutfuncdual}
  \comm{A}{B} =& \frac{1}{2} \ha A & \comm{D}{B} = &\frac{1}{2}
  \ha D,
\end{align}
and $ C $ is central \cite{HLR}.   It is interesting to note
that this is simply the undeformed algebra $ \heis^*$ with brackets as in \eqref{dualbrackets} and the simple identification
\begin{equation}
\begin{aligned}
\newJ^* &\simeq C, &&& \newS^* &\simeq B, &&& Z^* &\simeq A, &&& \bar{Z} &\simeq B,
\end{aligned}
\end{equation}
 One reason for the simplicity of this map may be the following simple relation between  the quantum $R$-matrix \eqref{eq:2} and the classical $r$-matrix \eqref{eq:rmatrix} in the representation \eqref{eq:matrices}
 \[
 R= 1\otimes 1 + \hbar r,\]
with {\em no} higher order corrections.

In order to complete the combinatorial quantisation  of Galilean gravity in 2+1 dimensions with cosmological constant one would need to complete the construction of the Newton-Hooke double, including the appropriate $*$-structures, and study unitary representations. The Hilbert space of Galilean quantum gravity  with cosmological constant can then, in principle,  be determined in terms of the representation theory of the Newton-Hooke double, see  \cite{Papageorgiou:2009zc} for  a review of the  details. However, even without knowing representations one can say something about the non-commutative spacetimes on which the Newton-Hooke double acts. The general procedure for computing this is given in \cite{MS}. As an algebra, the quantum Newton-Hooke spacetimes is again   $U_\ha(\heis) $, 
with the time generator $T$ being the central generator and the following commutator of the spatial coordinates $X_1$  and $X_2$
\begin{equation}
  \begin{aligned}
  \label{eq:nc_nh}
  [X_1,X_2] =&  8\pi G \hbar \frac{ e^{\sqrt{-\g{l}} T} -
    e^{-\sqrt{-\g{l}} T} }{2\sqrt{-\g{l}} } .
     \end{aligned}
\end{equation}
A  discussion of this algebra in  the limiting  case $\lambda=0$  can be found in \cite{Papageorgiou:2009zc}. We note that a spatial non-commutativity of this sort was also considered in in \cite{Daszkiewicz:2009px} in the context of a systematic study of twist deformations of the (unextended) Newton-Hooke algebra. Here we saw  that it arises naturally in (2+1)-dimensional quantum gravity with a cosmological constant.  
  The Newton-Hooke double acts on this non-commutative spacetime as a `quantised' isometry group. Formally, this is precisely the same structure as one encounters in the usual,  relativistic models of (2+1)-dimensional quantum gravity,   but working out the details remains a task for future work. It would also be interesting to see if the Galilean case with cosmological constant can be obtained by an appropriate contraction procedure from the relativistic treatment of the cosmological constant \cite{BNR}   in the quantum group setting. 


\section*{Acknowledgments}
GP acknowledges a PhD Scholarship by the Greek State Scholarship
Foundation (I.K.Y).

\end{document}